\documentclass[prb,twocolumn] {revtex4-1} 
\usepackage{graphicx}
\usepackage{url}
\usepackage{dcolumn}
\usepackage{upgreek}

\newcommand{\hnt}{\ensuremath h_{\mathrm{NIST\mbox{-}3}}}
\newcommand{\hne}{\ensuremath h_{\mathrm{NIST\mbox{-}2}}}
\newcommand{\hn}{\ensuremath h_{\mathrm{90}}}
\newcommand{\hs}{\ensuremath h_{\mathrm{07}}}
\newcommand{\hf}{\ensuremath h_{\mathrm{14}}}
\newcommand{\hfc}{\ensuremath h_{\mathrm{14corr}}}

\newcommand{\KJn}{\ensuremath K_{\mathrm{J-90}}}
\newcommand{\RKn}{\ensuremath R_{\mathrm{K-90}}}
\newcommand{\mymu}{{\ensuremath \upmu}}
\newcommand{\Js}{{\ensuremath \mbox{J\,s}}}

\begin{document}
\title{A summary of the Planck constant measurements using a watt balance with a superconducting solenoid at NIST}
\author{S Schlamminger$^1$\email{stephan.schlamminger@nist.gov}, R L Steiner$^1$, D Haddad$^2$, D B Newell$^1$, F Seifert$^2$, L S Chao$^1$, R Liu$^1$, E R Williams$^1$, J R Pratt$^1$ }
\address{
$^1$National Institute of Standards and Technology (NIST), 100 Bureau Drive Stop 8171, Gaithersburg, MD 20899, USA \\
$^2$Joint Quantum Institute, National Institute of Standards and Technology and University of Maryland, Gaithersburg, MD 20899, USA }

\begin{abstract}
Researchers at the National Institute of Standards and Technology have been using a watt balance, NIST-3, to measure the Planck constant $h$ for over ten years. Two recently published values disagree by more than one standard uncertainty. The motivation for the present manuscript is twofold. First, we correct the latest published number to take into account a recently discovered systematic error in mass dissemination at the Bureau International des Poids et Mesures (BIPM). Second, we provide guidance on how to combine the two numbers into one final result. In order to adequately reflect the discrepancy, we added an additional systematic uncertainty to the published uncertainty budgets. The final value of $h$ measured with NIST-3 is 
$h = 6.626\,069\,36(37)\times 10^{-34}\,\Js$. 
This result is $77(57) \times 10^{-9}$ fractionally higher than $h_{90}$. Each number in parentheses gives the value of the standard uncertainty in the last two digits of the respective value and $h_{90}$ is the conventional value of the Planck constant given by $h_{90}\equiv 4 /( \KJn^2\RKn)$, where $\KJn$ and $\RKn$ denote the conventional values of the Josephson and von Klitzing constants, respectively.
\end{abstract}

\maketitle

\section{Introduction}

Researchers at the National Institute of Standards and Technology (NIST) have published two results for the Planck constant, $\hs$ in 2007~\cite{rs07} and $\hf$ in 2014~\cite{ss14} using the same apparatus, the third generation of the NIST watt balance, referred to as NIST-3. The published values are
\begin{eqnarray}
\hs &=& 6.626\,068\,91(24)\times 10^{-34}\,\Js, \;\;\mbox{or} \nonumber \\
\frac{\hs}{\hn}-1&=& 8(36) \times 10^{-9}, \nonumber
\end{eqnarray}
and
\begin{eqnarray}
\hf &=& 6.626\,069\,79(30)\times 10^{-34}\,\Js, \;\;\mbox{or} \nonumber \\
\frac{\hf}{\hn}-1&=& 141(45) \times 10^{-9},\nonumber
\end{eqnarray}
respectively. Each number in parentheses gives the value of the standard uncertainty ($k=1$) in the last two digits of the quoted value. The symbol $h_{90}$ denotes the conventional value of the Planck constant given by $h_{90}\equiv 4 /( \KJn^2\RKn)$, where $\KJn$ and $\RKn$ denote the conventional values of the Josephson and von Klitzing constants, respectively.

Before we discuss the two measurements, we note that the 2014 result must be adjusted due to an offset in the SI unit of mass disseminated by the Bureau International des Poids et Mesures (BIPM). During the extraordinary comparison~\cite{extrao} of the international prototype of the kilogram (IPK) at the BIPM, it was found that the mass unit as maintained by the BIPM is 35\,$\mymu$g larger than the mass of IPK~\cite{mass_offset}. 

\begin{figure}[htb]
\centering
\includegraphics[width=0.95\columnwidth]{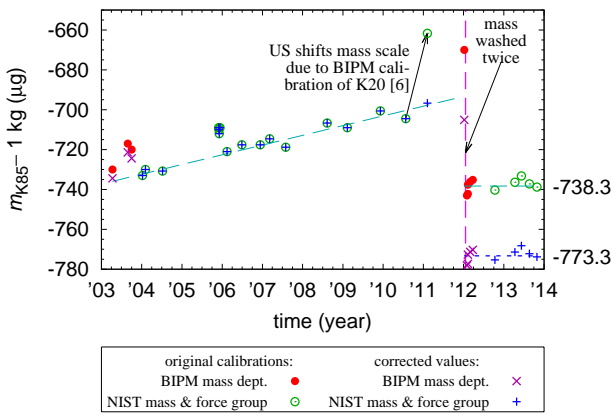}
\caption{The mass of the platinum-iridium prototype No. 85 (K85). The circles represent calibrations that were performed at the time of the experiment, either by the BIPM mass department (solid) or the NIST mass and force group (open). The diagonal crosses represent values that were calculated by applying a correction issued by the BIPM~\cite{BIPMcorr} to the original BIPM measurements. The upright crosses denote values that were calculated from measurements obtained by the NIST mass and force group by subtracting a correction. The correction is 0\,$\mu$g before December 2010 and  35\,$\mu$g after December 2010. The upright crosses represent the current best estimate of the mass of K85 relative to the IPK.}
\label{fig_mass}
\end{figure}

All measurements of the Planck constant discussed here were performed with the NIST platinum-iridium prototype No. 85, known as K85. Figure~\ref{fig_mass} shows the calibration history of K85. The first calibration certificate for K85 was issued by the BIPM in November 2003. Shortly after being put in service with NIST-3, the mass of this prototype drifted upwards with a rate of approximately 5\,$\mu$g per year. At the time of the NIST-3 experiment, this drift was taken into account in the calculations of the published values of $h$. In 2010, the United States shifted its mass scale by $45\,\mymu$g/kg, because it was found  during a routine calibration of the US national standard prototype K20 that the BIPM mass scale and the US mass scale differed by $45\,\mymu$g/kg~\cite{zk14}. At the end of 2011, K85 was sent for calibration to the BIPM, where it was washed two times.  The measurements of the mass of K85 performed by the NIST mass and force group and the BIPM in 2012 and 2013  are in agreement within uncertainties.  A constant value for the mass of K85 of $1\,\mbox{kg}-738.3\,\mymu$g was used to calibrate all NIST-3 measurements made in 2012 and 2013.

In 2014, during the extraordinary mass comparison, it was discovered that the mass unit as maintained by the BIPM is $35\,\mymu$g/kg larger than the mass scale set by the IPK~\cite{mass_offset}. It is believed that the difference between the mass unit as maintained by the BIPM and the IPK has accumulated between 2000 and 2014. Calibrations  made at the BIPM before 2000 are therefore deemed good calibrations with respect to the IPK~\cite{extrao}. But, calibration certificates issued by the BIPM after 2000 have to be corrected to be consistent with the international definition of the kilogram. According to a numerical model made by the BIPM, the calibration value assigned to K85 by the BIPM in 2003  must be decreased by approximately $4\,\mymu$g and the one assigned in 2012 by approximately $35\,\mymu$g~\cite{BIPMcorr}. The latter correction largely erases  the adjustment made to the US mass scale in 2010. 

After K85 had arrived at NIST at the end of 2003, it was measured against the national prototype, K20, by the NIST mass and force group. The national prototype, as well as the national check standard, K4, have been calibrated at the BIPM in 1999. At this time the mass unit as maintained by the BIPM was still synchronized to the IPK. Hence, it is believed that the calibration values determined by the NIST mass and force group for K85 from 2004 through the end of 2011 are accurate representations of the mass of K85 with respect to the mass of the IPK. The $h$ measurements taken during this time do not need to be corrected~\cite{zeina}.

In 2011, the mass and force group shifted its mass scale to be consistent with the mass scale given by the mass unit as maintained by the BIPM, which, as we now know, was different than the mass unit given by the IPK. Hence, all data taken with NIST-3 in 2012 and 2013 used a value for K85 that was tied to the mass scale as maintained by the BIPM, which was about $35\,\mymu$g~ different from the mass of the IPK at that time. Therefore, the measured value for $\hf$ needs to be corrected down relatively by $35\times 10^{-9}$. This yields
\begin{eqnarray}
\hfc &=&  6.626\,069\,56(30)\times 10^{-34}\,\Js, \;\;\mbox{or} \nonumber \\
\frac{\hfc}{\hn} - 1 &=&  106(45) \times 10^{-9}.\nonumber 
\end{eqnarray}

In summary, the time dependent shape of the correction to the NIST $h$ values is a step function rather than a smooth interpolation. The relative correction that must be applied to the published $h$ values is $0\times 10^{-9}$ for data taken before Dec. 2011 and   $-35\times 10^{-9}$ for data taken thereafter.

\begin{figure}[htb]
\centering
\includegraphics[width=0.95\columnwidth]{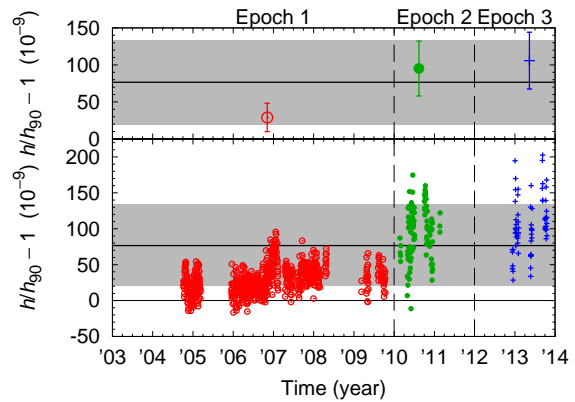}
\caption{Bottom Panel: All measurements obtained with NIST-3 using the  platinum-iridium prototype No. 85. Top Panel: The average value obtained during each epoch.  The error bars give the statistical standard deviation of the data. The data points in the third epoch in both panels are corrected down relatively by $35\times 10^{-9}$ from the original data to account for the offset between the IPK and the mass unit as maintained by BIPM that was discovered at the extraordinary comparison. The solid black lines in both panels give the unweighted average, $77\times 10^{-9}$, of the values from the three epochs listed in table~1. The gray bands around the solid lines indicate the one sigma uncertainty of the final value reported here.}
\label{fig1}
\end{figure}

The relative difference between the values of $\hs$ and $\hfc$ is $98  \times 10^{-9}$. All data obtained with NIST-3 using K85 are shown in Figure~\ref{fig1}. The data points after January 2012 are corrected by the  $35\times 10^{-9}$ as discussed above. The lower panel shows the data grouped in daily data runs.  The data can be divided into three epochs. The upper panel shows the values obtained by  taking the mean of the data in each epoch. The error bars indicate the statistical standard deviation of the data. The mean value and the statistical standard deviation for each epoch are also given in table~\ref{tab1}.

 The average value of the data in the first epoch is slightly larger than the value published in~\cite{rs07}. This is due to the fact that~\cite{rs07} includes data up to July 2006, while the first epoch includes data up to the end of 2009. The additional data increased the combined mean by a non-significant amount, i.e., the mean of the epoch and the published value are consistent within uncertainties.

\newlength{\myl}
\begin{table}[htb]
\begin{center}
\settowidth{\myl}{(42)}
\begin{tabular}{llrr}
epoch & time &  
\multicolumn{1}{c}{$\left(\frac{\displaystyle h}{\displaystyle\hn}-1\right)\times 10^9$} & 
\multicolumn{1}{c}{$\frac{\displaystyle \sigma_h}{\displaystyle h}\times 10^9$} \\
\hline
1 & 2004-2009  &29& 19\rule{0pt}{10pt} \\
2 & 2010/2011  &95 & 37\rule{0pt}{10pt} \\
3 & 2012/2013  &106&38\rule{0pt}{10pt} \\
\hline
\multicolumn{2}{c}{unweighted mean\rule{0pt}{10pt}} & 77 \\ 
\end{tabular}
\caption{The average values of $(h/\hn-1)\times 10^9$ for the three epochs. Here $\sigma_h$ denotes the statistical standard deviation in each epoch.}
\label{tab1}
\end{center}

\end{table}

The measured value of $h$ shifted by a relative amount of approximately $70\times 10^{-9}$, which at the time was not recognized, but we now term as the beginning of the second epoch of data recording. Nearly simultaneously with the (only later realized) long term shift, the day to day scatter of the data increased by a factor of two. At the time, the experiment was being incrementally improved while being studied to determine the cause of the sudden changes. Even recent reexamination of the change records and effects could find no clear, single cause for the shift or the increase in scatter.

Since 2010, the measured values were stable around a mean value of $h/\hn-1\approx 100\times 10^{-9}$ and the data could have been summarized into one longer epoch. However, significant changes were implemented starting in 2012 as summarized in~\cite{ss14}. Most importantly, the 2012 and 2013 measurements were performed blind with the researchers involved not knowing the true result until the unknown offset was publicly revealed in June 2013. Considering the fact that major changes between the second and third epoch did not alter the mean values gives us more confidence in the later number. Moreover, the shift in the mass scale discussed above can be interpreted as  a second blind bias added to the experiment. Yet, the 2012 and 2013 results agree with the measurements taken between 2010 and 2011.  Hence, we decided to give the later data more weight (two epochs after 2010 vs. one epoch before 2010). Weighting the data after 2010 more heavily, also reflects the fact that our understanding of the apparatus grew as a function of time. Towards the end of the data shown in Fig.~\ref{fig1}, we were more confident in our understanding of the system.

After more than fifteen years of experience with this apparatus, we believe that our best measurement of $h$ is obtained from the unweighted average of the values obtained in the three epochs listed in Table~1. This average  is $h/\hn-1=77\times 10^{-9}$.

Despite all our experience, we acknowledge that we do not understand the cause of the approximately $70\times 10^{-9}$ relative shift. This lack of understanding must be reflected in the uncertainty assigned to the final value. We assign half of the observed shift, $35\times 10^{-9}$, as an additional  relative uncertainty component to account for a possible unexplained systematic effect in the measurements. By adding this type B component in quadrature to the relative uncertainty published in~\cite{ss14}, a combined relative standard uncertainty of  $57\times 10^{-9}$ is obtained.

The final value for the Planck constant obtained with NIST-3 in ten years of measurement is 
\begin{eqnarray}
\hnt &=& 6.626\,069\,36(37)\times 10^{-34}\,\Js, \;\;\mbox{or}\nonumber\\
\frac{\hnt}{\hn} - 1 &=& 77(57) \times 10^{-9}.\nonumber
\end{eqnarray}


A previous version of the NIST watt balance (NIST-2) was used to determine a value of the Planck constant in a measurement campaign lasting four months in 1998~\cite{ew98,rs05}. The result
\begin{eqnarray}
\hne &=& 6.626\,068\,91(58)\times 10^{-34}\,\Js, \;\;\mbox{or} \nonumber \\
\frac{\hne}{\hn}-1&=& 8(87) \times 10^{-9} \nonumber
\end{eqnarray}
was obtained using two gold masses with a combined mass of 1\,kg. Hence, the value $\hne$ is independent of the mass changes discussed above and there is no need to modify this value or its uncertainty. Based on the previous estimate~\cite{CODATA06} of 0.14 for the correlation coefficient between $\hne$ and $\hs$, we estimate
\begin{eqnarray}
r(\hnt,\hne) & = & 0.09 \nonumber
\end{eqnarray}
to be an upper limit for the correlation coefficient between  $\hnt$ and $\hne$.

We thank the mass and force group at NIST, especially Zeina Kubarych and Patrick Abbott, for calibrating our masses and for sharing their expertise in mass metrology with us. 

\end{document}